\documentclass[twocolumn,prl,aps,amsfonts]{revtex4}

\def\<{\langle}
\def\>{\rangle}

\begin{document}
\title{Local indistinguishability: more nonlocality with less entanglement}

\author{Micha{\l} Horodecki, Aditi Sen(De), Ujjwal Sen,
and Karol Horodecki} 

\affiliation{Institute of
Theoretical Physics and Astrophysics, University of Gda\'{n}sk, 80-952
Gda\'{n}sk, Poland}

\begin{abstract}
We provide a first operational method for checking indistinguishability
of orthogonal states by local operations and classical communication (LOCC).
This method originates from the one introduced 
by Ghosh \emph{et al.} (Phys. Rev. Lett. \textbf{87}, 5807 (2001)), 
though we deal with {\it pure} states. 
We apply our method to show that an arbitrary complete 
multipartite orthogonal basis 
is indistinguishable by LOCC, if it contains at least one entangled state. 
We also show that probabilistic local distinguishing is possible for full basis
{\it if and only if} all vectors are product. We employ our method to 
prove local indistinguishability in an example with sets of pure states of \(3 \otimes 3\), which shows
 that one can have 
\emph{more nonlocality with less entanglement}, where ``more nonlocality'' is in the 
sense of ``increased local indistinguishability of orthogonal states''. This example also provides, to our knowledge,
 the only known example where \(d\) orthogonal states in \(d \otimes d\) are locally indistinguishable.

\end{abstract}
\maketitle
\vspace{-0.5cm}

Orthogonal quantum state vectors can always be distinguished if there 
are no restrictions to measurements that one can perform. If the vectors  are 
states of a system consisting of two distant subsystems, then 
there can be  natural restrictions for the measurements that can be done. 
In particular, if Alice and Bob (the parties holding the subsystems) 
cannot communicate quantum information, their possibilities 
significantly decrease.  Intuitively one feels that in such a case,
there will be a problem with distinguishing orthogonal entangled states,
while product ones should remain distinguishable. The first result 
in this area was rather surprising: in Ref. \cite{nlwe} 
the authors exhibited a set of orthogonal bipartite pure {\it product} states, 
that {\it cannot} be 
distinguished with certainty by local operations and classical communication (LOCC) 
\cite{huge}.
Another counterintuitive result was
obtained in Ref. \cite{Walgate}: {\it any} two orthogonal multipartite states
{\it can} be distinguished from each other by LOCC, irrespective of  
how entangled they are. The latter result was greatly 
extended in Refs. \cite{VSPM}. There is therefore a general question: which sets 
of orthogonal states are locally distinguishable?

To find that a given set is \emph{distinguishable} \cite{local}, one usually needs 
to build a suitable protocol. To show that the states are {\it not} 
distinguishable,
one can try to eliminate all possible measurements as 
in \cite{Walgate2}. Another way  is to employ somehow the theory of 
entanglement 
\cite{huge, Plenio, Vidal-mon, miaryQIC}. A typical statement proving 
such indistinguishability would be 
then: Alice and Bob cannot distinguish the states, as they would   
increase entanglement otherwise (which is impossible by LOCC).
The advantage 
of the latter method is that it allows to estimate the entanglement 
resources  needed to distinguish the states, that are non-distinguishable 
by LOCC. 

In Ref. \cite{TDL-hiding}, this approach was first used to check 
distinguishability between two mixed states (we will call it TDL method).  
Another powerful method  based on entanglement 
was recently  designed  in Ref. \cite{Aditi}  (we will call it 
GKRSS method). 
In this paper, building on those two concepts, we introduce first  
method that is \emph{operational}, i.e. it allows for systematic 
numerical checks. Moreover the method allows to obtain powerful 
analytic results. Our approach
provides a strong tool for investigation of distinguishability of 
sets of bipartite pure states, because it bases on deciding whether some pure state
can be transformed into some other pure states by LOCC,  the latter 
issue  being completely solved in a series of papers on entanglement 
measureses and entanglement manipulations with pure states 
\cite{Vidal-mon, Nielsen, Vidal, JP}.
Using it, we show that any full basis of  an
arbitrary number of systems is not 
distinguishable, if  at least one  of the vectors is entangled \cite{nieyeden}. For 
$2\otimes n$ systems it is then also ``only if", as product 
bases are distinguishable in this case \cite{UPB}. 
The result applies also to probabilistic distinguishability: 
we obtain  that a full basis is probabilistically distinguishable 
{\it if and only if} all vectors are product. 
As an illustration of the effectiveness 
of our presented method, 
we 
consider
an example of local indistinguishability
of an \emph{incomplete} basis 
which exhibits 
that it is possible to obtain \emph{more nonlocality with less entanglement}. 
To our knowledge, this is also the only known example of \(d\) indistinguishable states
in \(d \otimes d\).    


The application of entanglement 
theory to the problem of local distinguishability 
is not immediate. Imagine, that we want 
to distinguish between the four Bell states \cite{fourBell}. 
 If we were able to apply  by LOCC just the von Neumann 
measurement, then we could obviously create entanglement. 
Namely, if Alice and Bob start with any initial state (hence also possibly a disentangled one),
after  the von Neumann measurement, it collapses into one of Bell states.
This is of course impossible. We cannot however conclude at this 
moment,
that they are indistinguishable.  The clue is that we could 
distinguish between them, while destroying  them during the process. 
Thus Alice and Bob would get to know what state they shared, 
but the potential entanglement would be destroyed. This is 
actually the case in the Walgate \emph{et al.} protocol \cite{Walgate}, where 
one distinguishes between any two orthogonal (possibly) 
entangled states.

To employ entanglement theory in the distinguishability question, 
a more clever method should be applied. The general hint 
is to apply the measurement to some larger system. This concept is  
a basis for the TDL and GKRSS methods. In the first one 
\cite{TDL-hiding} the  authors  considered a state 
of four systems A, B, C, D:
\(
\psi=\psi_{AB}\otimes \psi_{CD}
\)
where $\psi_{AB}$ and $\psi_{CD}$ are maximally entangled states.
Then the measurement is applied to the AB part (cf. \cite{Lewenstein}). 
If the state 
after measurement is entangled, then one concludes that the measurement 
cannot be done  by use of LOCC, because entanglement can not be produced 
between the \(AB\) part 
and the \(CD\) part, without interaction between the two parts.

The GKRSS method   \cite{Aditi} is the following. Given 
the set of orthogonal states $\{\psi_i^{AB}\}_{i=1}^k$ to be distinguished, one 
builds a mixed state 
\begin{equation}
\label{fourparty}
\varrho =\sum_{i} {p_i}\left|\psi_i\right\rangle\left\langle\psi_i\right| \otimes 
\left|\phi_i\right\rangle\left\langle\phi_i\right|
\end{equation}
where $\phi_i$ are some entangled states of the CD system.
If Alice(A) and Bob(B)  are  able to distinguish between the states $\psi_i$ 
they can tell the result of their measurement to Claire(C) and Danny(D), who will 
then share states $\phi_i$ with probability $p_i$. One now compares 
the initial entanglement $E(\varrho)$ measured across the AC:BD cut 
and the final one  given by $\sum_i p_i E(\phi_i)$ according to any 
chosen entanglement measure $E$. 
If the states $\psi_i$ are  distinguishable by LOCC, then 
the final entanglement cannot be greater than the initial one; otherwise 
one could increase entanglement by LOCC \cite{monotone}. Thus, if 
we have 
\begin{equation}
\label{eq-aditi}
E(\varrho)< \sum_i p_i E(\phi_i)
\end{equation}
then the states $\psi_i$ are not distinguishable by LOCC. 
In Refs. \cite{Aditi, Aditi2} distillable entanglement was used as 
$E$.

Let us now  exhibit the method of the present Letter. It 
is a modification of the GKRSS method but an \emph{operational} one.
  Namely instead 
of classical correlations between AB and CD we will use 
quantum correlations. Consequently mixture (\ref{fourparty})
is replaced by the {\it superposition}
\begin{equation}
\psi_{ABCD}=\sum_i \sqrt{p_i}\left|{\psi}_{i}^{AB}\right\rangle\left|{\phi}_{i}^{CD}\right\rangle
\label{pure}
\end{equation}
The states $\phi_i$ will be  used here 
essentially to \emph{detect} as to whether a set of states are locally 
indistinguishable and as 
such we shall henceforth call them ``detectors".  At a first 
glance it seems that this approach should fail, because the pure state
is unlikely to have small entanglement. In \cite{Aditi} where 
mixtures are used, the possibility for the initial state $\varrho_{ABCD}$ 
to be separable in the AC:BD cut was much larger, as mixed states
are less coherent than pure ones; for a pure state to be separable,
it has to be product, while for mixed states, the very 
mixedness can decrease entanglement, or even produce 
separability \cite{volume}. Let us however exhibit 
the following example. Suppose that Alice and Bob are to distinguish between the Bell
 states  $\left|B_i\right\rangle$ \cite{fourBell}. As detectors, we take the same 
states (as in \cite{Aditi}). Our pure state is thus 
\begin{equation}
\label{Bell}
\left| \psi_{B} \right\rangle _{ABCD} =\frac{1}{2}\sum ^{4}_{i=1} \left| B_{i}\right\rangle _{AB}
\left| B_{i}\right\rangle _{CD} 
\end{equation}
One can see that this state can be written as 
\begin{equation}
\label{Michal}
\frac{1}{\sqrt{2}} \left(\left| 00\right\rangle + 
\left| 11\right\rangle\right)_{AC} \frac{1}{\sqrt{2}}
\left(\left| 00\right\rangle + \left| 11\right\rangle\right)_{BD} 
\end{equation}
So it turns out that it is \emph{product} in AC:BD cut, so that our 
method will work.  Assuming now the four Bell states to be locally 
distinguishable would immediately  imply that 
the state \( \left| \psi \right\rangle \) is entangled in the AC:BD 
cut which is the desired contradiction. This result was obtained in 
\cite{Aditi} and their  mixed state 
$\varrho_{ABCD}=1/4 \sum_{i} \left| B_{i} \right\rangle \left\langle B_{i} \right| \otimes 
\left|B_{i} \right\rangle \left\langle B_{i} \right|$ 
turned out 
to be separable in AC:BD (see also \cite{smolin}). 
Here we have a pure state which is product. 
Note that in this particular example, our method, even though 
originating from the GKRSS approach,  coincides with the TDL method.

The advantage of our approach over the GKRSS method is that 
for mixed states, it is usually hard to check the relation (\ref{eq-aditi})
for different entanglement measures. 
In our case we have pure states on both sides of the inequality, for which 
the set of all needed measures is known \cite{Vidal-mon, Vidal}.  
Even more:  Jonathan and Plenio \cite{JP},  generalizing the 
Nielsen result \cite{Nielsen},
have obtained a necessary and sufficient condition for the transformation 
from a pure state $\phi$ to an ensemble of pure states $\{p_i,\phi_i\}$.
The condition is efficiently computable. 
Namely, let $\lambda$ 
and $\lambda_i$  be vectors of the Schmidt coefficients
of $\phi$ and $\phi_i$ respectively.
Then the LOCC transition $\phi\to \{p_i,\phi_i\}$ is possible if and only if the 
vector $\sum_ip_i\lambda_i$ majorizes $\lambda$
\cite{majorization}. So our method consists of 
the following steps
\begin{itemize}
\item[(1)] Given the states $\{\psi_i^{AB}\}_{i=1}^k$ to be distinguished, 
choose $k$ detectors $\phi_i^{CD}$ and probabilities $p_i$.  
\item[(2)]Applying Jonathan-Plenio criterion \cite{JP}, check if the transition 
$\psi_{ABCD}\to \{p_i,\phi_i^{CD}\}$ is possible by LOCC 
(in AC:BD) where  $\psi_{ABCD}$ is of the form  (\ref{pure}).
\end{itemize}
If the transition is impossible,
the set of orthogonal states \(\left\{\psi_i\right\}_{i=1}^{k}\) are indistinguishable by LOCC.
The item (1) can be formulated more generally in the following way:
(1a) Choose $\psi_{ABCD}$ such that its reduction $\varrho_{AB}$ 
has the support spanned by $\psi_i^{AB}$'s;  
 (1b) Determine detectors $\phi_i^{CD}$ by writing 
$\psi_{ABCD}$ by means of $\psi_i^{AB}$. 
Let us mention here  that
we do not know of any example of a set of locally indistinguishable
orthogonal states whose local indistinguishability is in principle not obtainable by our method.

Now we will apply our method  to obtain the following proposition, 
where in fact we do not need an explicit use of the Jonathan-Plenio criterion.

{\bf Proposition.} Let $\psi_i^{AB}$ be a full orthogonal basis of an $m\otimes n$ 
system. Then we have: (1) If at least one of the vectors is entangled (see \cite{nieyeden}), 
the set cannot be perfectly distinguished by LOCC; (2) The set 
can be probabilistically distinguished if and only if all vectors are product. 

{\bf Remark.} We will not have ``if and only if" for item (1)
because there are orthogonal product bases that cannot be distinguished 
\cite{nlwe}. However item (1) would be ``only if" in \(2 \otimes n\),
as all product bases are locally distinguishable there \cite{UPB}. Note
also that item (2) \(\Rightarrow\) item (1).

{\bf Proof.} Consider the four party state  
\(
\left| \psi \right\rangle_{ABCD} = ( 1/ \sqrt {m}
\sum ^{m}_{i=1} \left| ii\right\rangle _{AC}) 
( 1/ \sqrt {n}\sum ^{n}_{j=1} 
\left| jj\right\rangle _{BD} ) 
\)
shared between Alice, Bob, Claire and Danny,
which is product across the AC:BD cut.
Written in AB:CD, this state takes the form 
\(1/ \sqrt {mn}\sum ^{mn}_{k=1} \left| k\right\rangle_{AB} \left| k\right\rangle_{CD}\). 

Let \( \left\{ \left| \psi_{1} \right\rangle, \left| \psi_{2} \right\rangle, \ldots,
 \left| \psi_{mn} \right\rangle \right\} \) 
be a set of \( mn \) orthonormal states of an \( m \otimes n \) system. We choose an unitary
operator 
\( U \) such that \( U \left|k\right\rangle = \left|\psi_{k}\right\rangle \) for all 
\(k = 1, 2, \ldots, mn \). We now use the \( U \otimes U^* \) invariance of the state 
\( \left|\psi\right\rangle \) in the AB:CD cut (see e.g. \cite{xor}) and write it as 
\(
1/ \sqrt {mn}\sum ^{mn}_{k=1} \left| \psi_{k}\right\rangle_{AB} \left| \psi_{k}\right\rangle^{*}_{CD} 
\),
where the complex conjugation is in the computational basis.

Therefore if Alice and Bob are able to locally distinguish between the \( \left|\psi_{k}\right\rangle \)s,
they could ring up Claire and Danny to tell which state they share, resulting in the creation of
the corresponding correlated state \( \left|\psi_{k}\right\rangle^{*} \) 
between Claire and Danny.

Now if at least one among the \( \left|\psi_{k}\right\rangle \)s is entangled, an assumption
of local distinguishability of the \( \left|\psi_{k}\right\rangle \)s would imply that the 
state \( \left|\psi\right\rangle \) has a nonzero amount of entanglement in the 
AC:BD cut \cite{key-6}. But this is 
forbidden as \( \left|\psi\right\rangle \) is product in the AC:BD cut.

Note that the above reasoning goes through irrespective of whether the local distinguishing 
protocol for the \( \left|\psi_{k}\right\rangle \)s is deterministic or 
probabilistic. This 
proves that an arbitrary complete set of orthogonal states of any bipartite system is locally
indistinguishable (deterministically as well as probabilistically) if 
at least one of the vectors is entangled. (Note that for the desired 
contradiction, the probabilistic protocol must have nonzero probability 
for at least one entangled state.) 

On the other hand, a given complete  product basis $\{v_i\}$
can be distinguished 
by von Neumann measurement $\sum_i |v_i\>\<v_i| (\cdot) |v_i\>\<v_i|$.
This is a separable operation  \cite{rains,Plenio} of the form 
$\sum_i A_i\otimes B_i (\cdot) A^\dagger_i \otimes B^\dagger_i$.
Such an operation can be probabilistically implemented by 
Alice and Bob \cite{Bennett} (it was 
first proven in \cite{Lewenstein}):
they pick random $i$, and probabilistically perform operation 
$A_i\otimes B_i(\cdot) A^\dagger_i \otimes B^\dagger_i$. \(\Box\)










\textbf{Generalisation of the proposition.} 
In \( d_{1} \otimes d_{2} \otimes \ldots  \otimes d_{N}\), 
a full orthogonal basis can be distinguished probabilistically 
if and only if all vectors are product (i.e., of the form
\(\left|\eta_1\right\rangle \otimes
 \left|\eta_2\right\rangle \otimes \ldots \otimes\left|\eta_N\right\rangle\)) \cite{notgenuineN}.

The ``only if" part of the generalised proposition is immediate,
from the Proposition for the bipartite case, once we 
note that a multiparty entangled state must 
be entangled in at least one bipartite cut. Note also
that if a set of multipartite states is indistinguishable 
in a bipartite cut, it would obviously remain so, if we lessen 
the allowed set of operations by restricting the parties 
within one cut to remain at distant locations.  Since multipartite 
separable maps can be preformed probabilistically (the same 
reasoning as above), we obtain also the ``if" part.
Note that our presented method for testing local 
indistinguishability of a set of \emph{bipartite} orthogonal states
cannot be extended in its full generality to the multipartite situation
 as the 
Jonathan-Plenio criterion \cite{JP} has not been as yet generalised to more 
than two parties. 




To see the effectiveness of the presented method,  
we apply it to obtain an interesting example of indistinguishability of an 
\emph{incomplete} basis of orthogonal states. 
First, note that the set \(S\) consisting of the following 
maximally entangled states (without normalisation)
 in \(3 \otimes 3\) are distinguishable locally:
\begin{equation}
\begin{array}{rcl}
\label{max3x3}
\psi_{1}=\left|00\right\rangle + \omega \left|11\right\rangle &+&  \omega^{2} \left|22\right\rangle, 
\psi_{2}=\left|00\right\rangle + \omega^{2} \left|11\right\rangle + \omega \left|22\right\rangle,  \\
\psi_{3}&=&\left|01\right\rangle + \left|12\right\rangle + \left|20\right\rangle.
\end{array}
\end{equation}
(\(\omega\) is a nonreal cube root of unity.) 
The set \(S\) can be distinguished locally by making a projective measurement (on any one side)
in the basis
\(\{1/\sqrt{3}\left(\left|0\right\rangle+\left|1\right\rangle+\left|2\right\rangle\right)\),
\(1/\sqrt{3}\left(\left|0\right\rangle+\omega\left|1\right\rangle+\omega^{2}\left|2\right\rangle\right)\),
\(1/\sqrt{3}\left(\left|0\right\rangle+\omega^{2}\left|1\right\rangle+\omega\left|2\right\rangle\right)\}\)
and a subsequent classical communication to the 
other party (see also \cite{Dong}). 

Having shown this, what would be our expectation 
for the set of states containing the same states as in \(S\) but for the last state
\(\left|\psi_{3}\right\rangle\),
which is replaced by a \emph{product} state \(\psi_3^{'} =\left|01\right\rangle\)? 
The above Propositions seem to indicate that as we put more and more entanglement into 
the system, the system tends to become locally indistinguishable. This is also
the expectation obtained from the recent work of Walgate and Hardy \cite{Walgate2}. 
But one can check by taking \(B_{i}\)s \((i=1, 2, 3)\) as detectors and with probabilities
\(p_i\) as \(\left(.16, .16, .68\right)\), that 
the transition 
\(\sum_{i=1}^{2}\sqrt{p_{i}}\left|\psi_{i}\right\rangle_{AB}\left|B_{i}\right\rangle_{CD}
+ \sqrt{p_{3}}\left|\psi_{3}^{'}\right\rangle_{AB}\left|B_{3}\right\rangle_{CD}\)
\(\rightarrow \left\{p_{i}, \left|B_i\right\rangle_{CD} \right\}\) is forbidden by the 
Jonathan-Plenio criterion \cite{JP}. 
Consequently the set \(S^{'}\), containing the states (without normalisation)
\begin{equation}
\begin{array}{rcl}
\label{min3x3}
\psi_{1}= \left|00\right\rangle + \omega \left|11\right\rangle +  & \omega^{2} & \left|22\right\rangle, 
\psi_{2} =\left|00\right\rangle + \omega^{2} \left|11\right\rangle + \omega \left|22\right\rangle, \\
\psi^{'}_{3}&=& \left|01\right\rangle
\end{array}
\end{equation}
is \emph{indistinguishable} by LOCC \cite{conjecture}.
This simple 
example shows that the intuition 
that we tried to obtain 
from our Propositions as well as from the work of 
Walgate and Hardy \cite{Walgate2} is not true.
Reducing entanglement from 
the system can in fact \emph{increase} the nonlocality 
of the system. 
This may therefore further the process of ``disentangling" nonlocality (in the sense 
of local indistinguishability) from entanglement \cite{nlwe, Walgate, UPB, VSPM}.
Note that, to our knowledge, 
this is the only known example of a set of \(d\) indistinguishable states in \(d \otimes d\).

Since our method is based on entanglement measures \cite{monotone},
 there is a question, whether all 
operations that  cannot be performed by LOCC would increase at least 
one entanglement measure. Most likely it is the case, 
i.e. the set of LOCC doable operations is described by the 
set of  entanglement measures.

To conclude, we provide a powerful method allowing for efficient investigation of indistinguishability of 
orthogonal vectors via LOCC. 
We were able to prove general statements for indistinguishability of full bases as well as 
to provide a counterintuitive example.
The question arises whether our method gives  the if and only if criterion. In other words,
 given an ensemble, is  it true that they are indistinguishable by LOCC if and only if
we can find such detectors so that our method will detect indistinguishability of the ensemble? 
For example, there exist sets of   product  states
that  can be distinguished by
separable  operations \cite{rains} but not by LOCC \cite{nlwe, UPB, UPB1}. 
Can our method detect such cases? If the answer is ``yes'', it would imply that there is
an entanglement measure  that  {\it can } increase  under separable operations
(even though it of course cannot increase
under LOCC). In our method we go from pure states to pure states, and the set of entanglement measures that are
 responsible for such possibility is well
known and finite \cite{Vidal-mon,JP}. They are sums of squares of $k$ largest Schmidt coefficients ($k=1,\ldots,d$,
 where $d$ is
 dimension of subsystem). There remains an open question as to whether they could increase 
under separable operations. If the answer is ``yes'', then our method can be applied to analyse 
distinguishability of aforementioned product states.
It is however clear that we could not then apply our method with the initial 
state as product with respect to AC:BD cut. This is because
 separable operations cannot produce entangled state out of product ones, 
but can distinguish between the states of interest.

We thank C.H. Bennett for 
a discussion at the European 
Research 
Conference 
on 
Quantum Information in San Feliu de Guixols, 2002
and L. Hardy for a discussion at
the International Conference on Quantum Information in Oviedo, 2002.
We also thank S. Ghosh, G. Kar, J. Oppenheim, A. Roy, 
D. Sarkar, B. Synak  and J. Walgate. 
The work is supported by the European Community under project EQUIP,
Contract No. IST-11053-1999 and by the University of Gda\'{n}sk, Grant No. BW/5400-5-0236-2.


\begin{thebibliography}{10}
\bibitem{nlwe}
C.H. Bennett, D.P. DiVincenzo, C.A. Fuchs, T. Mor, E. Rains, 
P.W. Shor, J.A. Smolin, and W.K. Wootters, Phys. Rev. A
\textbf{59}, 1070 (1999).
\bibitem{huge} C. H. Bennett, D. P. DiVincenzo, J. A. Smolin, and 
W. K. Wootters, Phys. Rev. A {\bf 54}, 3824 (1996). 
\bibitem{Walgate}
J. Walgate, A.J. Short, L. Hardy, and V. Vedral, Phys. Rev. Lett. {\bf 85}, 4972 (2000).
\bibitem{VSPM} S. Virmani, M.F. Sacchi, M.B. Plenio, and D. Markham, 
Phys. Lett. A \textbf{288}, 62 (2001); Y.-X. Chen and D. Yang, Phys. Rev. A \textbf{64}, 064303 (2001); \emph{ibid.}
\textbf{65}, 022320 (2002).
\bibitem{local} In this Letter, ``distinguishable'' (``indistinguishable'') means ``locally distinguishable'' 
(``locally indistinguishable''). And we only consider indistinguishability 
of \emph{orthogonal} states.
\bibitem{Walgate2}
J. Walgate and L. Hardy, Phys. Rev. Lett. \textbf{89}, 147901 (2002).
\bibitem{Plenio} V. Vedral, M.B. Plenio, M.A. Rippin, and P.L. Knight, 
Phys. Rev. Lett. \textbf{78},  2275 (1997); 
V. Vedral and M.B. Plenio, Phys. Rev.  A {\bf 57}, 1619 (1998). 
\bibitem{Vidal-mon} 
G. Vidal,  J. Mod. Opt. {\bf 47}, 355 (2000).  
\bibitem{miaryQIC}
M. Horodecki, Q. Comp. Inf. {\bf 1}, 7 (2001). 
\bibitem{TDL-hiding}
B.M. Terhal, D.P. DiVincenzo, and D.W. Leung, 
Phys. Rev. Lett. {\bf 86}, 5807 (2001).
\bibitem{Aditi} S. Ghosh, G. Kar, A. Roy, A. Sen(De), and U. Sen,
Phys. Rev. Lett. \textbf{87}, 277902 (2001). 
\bibitem{Nielsen}
M.A. Nielsen, Phys. Rev. Lett. {\bf 83}, 436 (1999). 
\bibitem{Vidal} G. Vidal, Phys. Rev. 
Lett. {\bf 83}, 1046 (1999).   
\bibitem{JP} D. Jonathan and M.B. Plenio, Phys. Rev. Lett. \textbf{83}, 1455
(1999).  
\bibitem{nieyeden} Note that it is not possible to have a complete basis in a tensor product
Hilbert space where all but \emph{one} states are product 
 \cite{UPB}. 
\bibitem{UPB} C.H. Bennett, D.P. DiVincenzo, 
T. Mor, P.W. Shor, J.A. Smolin, and B.M. Terhal, Phys. Rev. Lett.
\textbf{82}, 5385 (1999).  
\bibitem{Lewenstein}
J.I. Cirac, W. D\"ur, B. Kraus, and M. Lewenstein, 
Phys. Rev. Lett. {\bf  86}, 544  (2001).
\bibitem{fourBell} The four Bell states are 
\(\left| B_{1,2}\right\rangle  =  1/\sqrt{2}\left( \left| 00\right\rangle \pm \left| 11\right\rangle \right)\)
and 
\(\left| B_{3,4}\right\rangle   =  1/\sqrt{2}\left( \left| 01\right\rangle \pm \left| 10\right\rangle \right)\).
\bibitem{monotone} In the Vidal approach \cite{Vidal-mon}, entanglement (or entanglement 
measure) is defined simply
as a quantity that does not increase under LOCC (cf. \cite{RP}).
\bibitem{RP}
S. Popescu and D. Rohrlich, Phys. Rev. A \textbf{56}, R3319 (1997).
\bibitem{Aditi2} S. Ghosh, G. Kar, A. Roy, D. Sarkar, A. Sen(De), 
and U. Sen, Phys. Rev. A \textbf{65}, 062307 (2002). 
\bibitem{volume}
K. \.Zyczkowski, P. Horodecki, A. Sanpera, and M. Lewenstein,
Phys. Rev. A {\bf 58},  883 (1998);
K. \.Zyczkowski,  Phys. Rev. A {\bf 60}, 3496 (1999). 
\bibitem{smolin} J.A. Smolin, Phys. Rev. A \textbf{63}, 032306 (2001). 
\bibitem{majorization}
If $x=(x_1,\ldots, x_d)$ and  $y=(y_1,\ldots, y_d)$ are real 
$d$ dimensional vectors,
$x \prec y$ ($x$ is majorized by $y$) if 
\(\sum_j^k x_{j}^{\downarrow} \leq \sum_j^k y_{j}^{\downarrow}
\quad \forall k \in \left\{1,\ldots, d\right\}\), where \(x_{j}^{\downarrow}\) and 
\(y_{j}^{\downarrow}\) are 
elements of \(x\) and \(y\), 
set 
in
decreasing order.
\bibitem{xor}
M. Horodecki and P. Horodecki, Phys. Rev. A {\bf 59}, 4206 (1999). 
\bibitem{key-6} Note that \(\left|\phi\right\rangle^{*}\) is 
entangled whenever 
\(\left|\phi\right\rangle\) is entangled.
\bibitem{Bennett} There is a direct, geometrical argument showing that a complete 
product basis is 
always probabilistically distinguishable  [C.H. Bennett, private communication].
\bibitem{notgenuineN}The entangled state needed for the indistinguishability in the statement of the 
generalised proposition may not be a genuine \(N\)-party entanglement. 
E.g., in the \(3\)-qubit case, even a state \((a\left|00\right\rangle + b\left|11\right\rangle) \otimes 
\left|0\right\rangle\) in a complete orthogonal basis would be sufficient for local indistinguishability
of the basis.
\bibitem{Dong} Y.-X. Chen and D. Yang, quant-ph/0204152. 
\bibitem{conjecture} 
The three states 
\(\psi_1\), \(\psi_2\) and \(a\left|01\right\rangle + b\left|12\right\rangle + 
c\left|20\right\rangle\) are indistinguishable for a continuous region 
around \((a, b, c) = (1, 0, 0)\)
for the same detectors and probabilities. We are willing to {\it conjecture}
that \(\psi_1\), \(\psi_2\) and \(a\left|01\right\rangle + b\left|12\right\rangle + 
c\left|20\right\rangle\) are locally indistinguishable 
for the whole range of \((a, b, c)\) except when the last is maximally entangled.
\bibitem{orNielsen} Actually in the considered case, all final states 
have the same $\lambda_i$, so that the Jonathan-Plenio criterion 
reduces to Nielsen's one \cite{Nielsen}.
\bibitem{rains} E.M. Rains, Phys. Rev. A {\bf 60}, 173 (1999). 
\bibitem{UPB1} D.P. DiVincenzo, T. Mor,
P.W. Shor, J.A. Smolin, and B.M. Terhal, quant-ph/9908070.
\bibitem{peres} A. Peres, Phys. Rev. Lett. \textbf{77}, 1413 (1996).


\end{thebibliography}
\end{document}